\documentstyle[aps,prl,multicol]{revtex}
\begin{document}
\title{Exotic ground states and impurities in multiband superconductors}
\author{D.F. Agterberg$^{\dagger}$, Victor Barzykin$^{\dagger}$, 
and Lev P. Gor'kov$^{\dagger,*}$}
\address{$^{\dagger}$National High Magnetic Field Laboratory, 
Florida State University,\\
1800 E. Paul Dirac Dr., Tallahassee, Florida 32310 \\
and \\
$^*$L.D. Landau Institute for Theoretical Physics,
Chernogolovka, 142432, Russia
}
\maketitle
\begin{abstract}
We consider the effect of isotropic impurity scattering  on the exotic
superconducting states
that arise from
the usual BCS mechanism in substances of cubic and hexagonal symmetry
where the Fermi surface contains inequivalent but degenerate pockets
({\it e.g.} 
 around several points of high symmetry). As examples we look at
CeCo$_2$, CeRu$_2$, and
LaB$_6$;
all of which have such Fermi surface topologies and the former exhibits
unconventional
superconducting behavior.  
We find that while these non $s$-wave states are suppressed
by non-magnetic impurities, the suppression is much weaker than
would be expected for unconventional superconductors with isotropic
non-magnetic impurity scattering.
\end{abstract}

Recently it was shown that in substances with cubic or hexagonal
symmetry, exotic superconducting states 
can arise from conventional BCS
mechanisms \cite{ABG99}. This can occur when these metals have 
Fermi surface (FS) pockets 
that are centered at  
or around some three-fold or higher degenerate 
symmetry points of the Brillouin zone (BZ).  
The resulting superconducting state is  either a conventional
$s$-wave state or corresponds to a multidimensional superconducting
representation for which the ground state breaks time reversal symmetry. 
Such FS topologies exist in many superconductors. 
Examples are CeRu$_2$ \cite{hed95}, LaB$_6$ \cite{ark75}, and CeCo$_2$
\cite{sug95}; 
the latter exhibits  a  low temperature behavior specific heat that
appears to vary as
a power law  with temperature \cite{sug95,aok97} which is  
consistent with unconventional
superconductivity.   
Here we investigate the stability of these exotic superconducting states
in the presence of non-magnetic isotropic impurity scattering.  
In particular we examine in detail a model that applies both to CeCo$_2$
and
to the FS pockets centered at the six N-points of a body-centered cubic 
(BCC) lattice and discuss the
results of the analogous theory  for CeRu$_2$ and LaB$_6$. 
We show that the exotic superconducting  states are less susceptible to
non-magnetic 
impurities than might be expected from earlier theories of
isotropic impurity scattering  in single band unconventional
superconductors 
\cite{gor85,ued85}. 

A BCS approximation for the multi-band case 
 will be used 
(see {\it e.g.} \cite{suh59}). The form 
of the interaction parameters describing the two electron scattering
on and between the different degenerate pockets of the FS
 is fixed by
symmetry. Consequently, the resulting superconducting state need not
be $s$-wave when the interaction does not depend on the direction of the
Fermi 
momenta of an individual pocket.
In our earlier work \cite{ABG99} we considered three cases in detail
a) three FS pockets centered about the
X-points of a simple cubic lattice;
b) three FS pockets at the M-points of the hexagonal
lattice; c) four FS pockets at the L-points in the face-centered cubic
(FCC)
lattice. Case a) may describe the superconducting states in LaB$_6$ and
in CeRu$_2$ 
since these materials both have FS pockets centered at the X-point.   

It has recently been demonstrated that the specific heat exhibits 
a power law temperature dependence at low temperatures in CeCo$_2$ 
\cite{sug95,aok97}.
For this reason  
we consider here FS pockets centered at the six 
N-points of the 
BCC lattice which is formally identical to
that in FCC CeCo$_2$ for the FS pockets 
which are located along the $(1,1,0)$
and equivalent directions \cite{sug95,sug96}
(note that in this material
there also exist FS sheets centered at the $\Gamma$ point).  

The Hamiltonian for several separate pieces of the FS can be
written in the following form:
\begin{equation}
H= \sum_{\alpha \sigma {\bf p}} \epsilon({\bf p})
a^{\dagger}_{\alpha \sigma}({\bf p}) a_{\alpha \sigma}({\bf p})
+ {1 \over 2} \sum_{{\bf k},{\bf k'},{\bf q}}
\sum_{\alpha \beta \sigma \sigma'} \lambda_{\alpha \beta}({\bf q})
a^{\dagger}_{\alpha \sigma}({\bf k+q}) a^{\dagger}_{\beta \sigma'}({\bf
k'-q})
a_{\alpha \sigma'}({\bf k'}) a_{\beta \sigma}({\bf k}),
\end{equation}
\noindent where  $\sigma$ and $\sigma'$ are
spin indices, $\lambda_{\alpha \beta}({\bf q})$ includes the interaction
for scattering two electrons from the pocket $\alpha$ into pocket
$\beta$
which is due to both Coulomb and electron-phonon terms.
For simplicity we will take $\lambda_{\alpha,\beta}({\bf q})$ to be
independent
of the direction of ${\bf q}$.
Scattering by impurities  is described by the Hamiltonian 
\begin{equation}
H_{imp}=\sum_{\sigma}\sum_{k,q}\sum_{\alpha,\beta}\sum_j
\exp[i({\bf q}-{\bf k}){\bf R}_j]V_{\alpha,\beta}({\bf k},{\bf q})
a^{\dagger}_{\beta \sigma}({\bf k})a_{\alpha \sigma}({\bf q}).
\end{equation}
Introducing the anomalous Green's function
$\hat{\cal{F}}_{\alpha}(x-x')$ for each FS sheet $\alpha$,
the corresponding Gor'kov equations can be found  
for the case of singlet pairing in exactly the same 
manner as in Ref~\cite{AGD}. 
Including the
isotropic impurity scattering within the Born approximation  
the following Gor'kov equations are found 
\begin{eqnarray}
[i\omega_n-\epsilon_{\alpha}({\bf k})-\bar{{\cal G}}_{\alpha}(\omega)]
{\cal G}_{\alpha}({\bf k},i\omega_n)
-[\Delta_{\alpha}^*+\bar{{\cal F}_{\alpha}^{\dagger}}(\omega)]
{\cal F}_{\alpha}({\bf k},i\omega_n)=1
\nonumber \cr
[i\omega_n+\epsilon_{\alpha}({\bf k})+\bar{{\cal G}}_{\alpha}(-\omega)]
{\cal F}_{\alpha}^{\dagger}({\bf k},i\omega_n)
-[\Delta_{\alpha}^*+\bar{{\cal F}^{\dagger}_{\alpha}}(\omega)]
{\cal G}_{\alpha}({\bf k},i\omega_n)=0
\end{eqnarray}
where 
\begin{eqnarray}
\bar{{\cal G}}_{\alpha}(\omega)=n_i\sum_{{\bf k},\beta}
|u_{\alpha,\beta}|^2
{\cal G}_{\beta}({\bf k},i\omega_n)\nonumber \cr
\bar{{\cal F}}_{\alpha}(\omega)=n_i\sum_{{\bf k},\beta}
|u_{\alpha,\beta}|^2
{\cal F}_{\beta}({\bf k},i\omega_n) \nonumber \cr
\Delta^*_{\alpha} = - T \sum_{{\bf k},\beta}\sum_n 
\lambda_{\beta \alpha}{\cal F}^{\dagger}_{\beta}(\omega_n, {\bf k})
\end{eqnarray}
$|u_{\alpha,\beta}|^2= 
\int \frac{d\Omega}{4\pi}\int \frac{d\Omega^{\prime}}{4\pi}
|V_{\alpha,\beta}({\bf k}_F,{\bf k}^{\prime}_F)|^2$ and $n_i$ is 
the concentration of impurities. 
Introducing $\tilde{\Delta}_{\alpha,n}=\Delta_{\alpha}+\bar{{\cal
F}^{\dagger}_{\alpha}}(\omega)$
and $i\tilde{\omega}_{\alpha,n}=i\omega_n-\bar{{\cal
G}}_{\alpha}(\omega)$, 
the gap 
and self-energy equations can be expressed as 
\begin{equation}
\Delta_{\alpha}=T\pi\sum_{\beta,n}\frac{N_0\lambda_{\alpha,\beta}
\tilde{\Delta}_{\beta,n}}
{\sqrt{\tilde{w}_{n,\beta}^2+|\tilde{\Delta}_{\beta,n}|^2}},
\label{gap1}
\end{equation}
\begin{equation}
\tilde{w}_{n,\alpha}=w_n+\sum_{\beta}\frac{\Gamma_{\alpha,\beta}
\tilde{w}_{n,\beta}}{\sqrt{\tilde{w}_{n,\beta}^2+|\tilde{\Delta}_{\beta}|^2}}
\label{gap2}, 
\end{equation}
and
\begin{equation}
\tilde{\Delta}_{n,\alpha}=\Delta_\alpha+\sum_{\beta}\frac{\Gamma_{\alpha,\beta}
\tilde{\Delta}_{n,\beta}}{\sqrt{\tilde{w}_{n,\beta}^2+
|\tilde{\Delta}_{\beta,n}|^2}}
\label{gap3}
\end{equation}
where 
$\Gamma_{\alpha,\beta}=\pi N_0 |u_{\alpha,\beta}|^2$, and 
$N_0$ is the normal density of states on a single pocket.
Equations (\ref{gap1}, \ref{gap2}),
and (\ref{gap3})
have been studied in the context of multi-band superconductivity by a
variety
of authors (see, {\it e.g.}, \cite{mos66,sch77,gol97}).   

We consider here FS pockets centered at the six N points of the BCC 
lattice. This situation is formally identical to that describing 
the FS pockets in FCC CeCo$_2$ 
located along the (110) and equivalent directions.  
The N points of the BCC lattice lie at $\frac{1}{2}{\bf b}_i$ and 
$\frac{1}{2}({\bf b}_i-{\bf b}_j)$ with $i>j$ where the
${\bf b}_i$ are the reciprocal lattice basis vectors 
[${\bf b}_1=\frac{2\pi}{a}(0,1,1)$, ${\bf b}_2=\frac{2\pi}{a}(1,0,1)$,
and
${\bf b}_3=\frac{2\pi}{a}(1,1,0)$]. The interaction between the
electrons 
forming a Cooper pair 
on Fermi pockets centered at these points takes the form
\begin{equation}
V=\pmatrix{\lambda & \nu & \mu & \nu & \nu & \nu \cr
\nu & \lambda & \nu & \mu & \nu & \nu \cr 
\mu & \nu &\lambda & \nu & \nu & \nu \cr
\nu & \mu & \nu & \lambda & \nu &  \nu \cr
\nu & \nu & \nu & \nu & \lambda & \mu \cr
\nu & \nu & \nu & \nu & \mu & \lambda }
\end{equation}
where $\lambda$ is the interaction on the same pocket, $\mu$
couples two pockets connected by a  $\frac{4\pi}{a}(1,0,0)$, 
$\frac{4\pi}{a}(0,1,0)$, or a $\frac{4\pi}{a}(0,0,1)$  translation, and
$\nu$ characterizes the remaining couplings between nearest neighbor N
points. 
The matrix for the 
impurity scattering (with
elements $\Gamma_{\alpha,\beta}$) has the same form and will
be described by elements $\Gamma_0,\Gamma_1$, and $\Gamma_2$ 
replacing $\lambda$, $\nu$, and $\mu$ respectively. 

Consider the clean limit,   
the linearized gap equation is 
\begin{equation}
\Delta_{\alpha}^*  = - \sum_{\beta} N_0 \lambda_{\alpha \beta}
\Delta_{\beta}^* \ln\left({2 \gamma \omega_D \over \pi T_c}\right).
\label{Tcc}
\end{equation}
The six 
$\Delta_{\alpha}$ transform among  each other under cubic symmetry 
transformations
forming a 6D reducible representation of the cubic
group $O_h$, which is split into a $1D$ $A_{1g}$, 
a $2D$ $E_g$, and a $3D$ $F_{2g}$  irreducible representation.
These three representations correspond to different order
parameters with three critical temperatures:
\begin{eqnarray}
T_{c0,A} &=& {2 \gamma \omega_D \over \pi}
\exp\left[ \frac{1}{N_0(\lambda + \mu +4\nu)}\right] \\
T_{c0,E} &=& {2 \gamma \omega_D \over \pi}
\exp\left[\frac{1}{N_0(\lambda + \mu-2\nu)}\right]  \\
T_{c0,F} &=& {2 \gamma \omega_D \over \pi}
\exp\left[\frac{1}{N_0(\lambda -\mu)}\right]
\label{TC}
\end{eqnarray}
where the factors in the exponentials must be negative for a non-zero
transition temperature. 
When $\nu>0$ or $2\nu+\mu>0$ ({\it i.e.} for repulsive
inter-pocket interactions) the higher dimensional 
representations have the higher $T_c$. 
The basis wave function for 1D $A_{1g}$ identical representation is
\begin{equation}
l = (\Delta_1 + \Delta_2 + \Delta_3 +\Delta_4 + \Delta_5
+\Delta_6)/\sqrt{6},
\label{1Dr}
\end{equation}
the basis wave functions for the 2D $E_g$  representation can be chosen
as
\begin{eqnarray}
\eta_1 &=& (\Delta_1 + \epsilon \Delta_2 + \Delta_3+ \epsilon \Delta_4 +
\epsilon^2 \Delta_5 +\epsilon^2\Delta_6)/
\sqrt{6} \nonumber\\
\eta_2 &=& (\Delta_1 + \epsilon^2 \Delta_2 + \Delta_3+ \epsilon^2
\Delta_4 +
\epsilon \Delta_5 +\epsilon\Delta_6)/\sqrt{6}
\label{2Dr}
\end{eqnarray}
where $\epsilon = \exp(2 \pi i/3)$, 
and the basis wave functions for the 3D $F_{2g}$ representation can 
be chosen as
\begin{eqnarray}
\eta_x &=& (\Delta_5-\Delta_6)/\sqrt{2} \nonumber \\
\eta_y &=& (\Delta_1  - \Delta_3) /\sqrt{2} \ \ \ \ \ \label{3Dr} \\
\eta_z &=& (\Delta_2 - \Delta_4)/\sqrt{2} \nonumber.
\end{eqnarray}
Following Refs. \cite{gor59,ABG99}, the Ginzburg Landau functional for
the 
2D and 3D representations are 
found to be
\begin{equation}
\delta F_{E}/N_0 = {T-T_{c0,E} \over T_{c0,E}} (|\eta_1|^2+|\eta_2|^2)+
{7 \zeta(3) \over 96 \pi^2 T_{c0,E}^2}(|\eta_1|^4 + |\eta_2|^4+ 4 
|\eta_1|^2 |\eta_2|^2)
\label{F2D}
\end{equation}
\begin{equation}
\delta F_{F}/N_0 = {T-T_{c0,F} \over T_{c0,F}} (\vec{\eta}
\vec{\eta}
^*) +
{7 \zeta(3) \over 32 \pi^2 T_{c0,F}^2}
(|\eta_x|^4 + |\eta_y|^4+ |\eta_z|^4).
\label{F3D}
\end{equation}
Eq.~(\ref{F2D}) for the $E_g$ representation implies  
$(\eta_1,\eta_2)=(1,0)$ is a stable ground state.
Following the notation of Ref.~\cite{gor87}
this corresponds to the superconducting class 
$O(D_2)$. The properties of such a ground state are the same as
those found for the analogous calculation for FS pockets centered at 
the three X points in a simple cubic lattice and this is also the
situation that
applies to both LaB$_6$ and CeRu$_2$  (we refer to Ref.~\cite{ABG99}
for more details). Eq.~(\ref{F3D}) for the $F_{2g}$ representation  
implies the ground state 
solution $(\eta_x,\eta_y,\eta_z)=[1,\exp(i\phi_1),\exp(i\phi_2)]$ 
where $\phi_1$ and $\phi_2$ are arbitrary phase factors.
The degeneracy arising from the phases $\phi_1$ and $\phi_2$ is 
an artifact of the BCS theory and is not lifted by higher order terms
in free energy functional. In the notation of Ref.~\cite{gor87} the 
free energy of Eq.~(\ref{F3D}) places the system on boundary of two
phases with superconducting classes 
$D_3\times R$ and $D_3(E)$. A similar situation arose for FS pockets
centered at the four L points of a face centered cubic lattice 
where a degeneracy was found between solutions corresponding
to the classes $D_4^{(2)}(D_2)\times R$ and $D_4(E)$ of the $F_{2g}$
representation \cite{ABG99}.  
The presence of a FS centered at the $\Gamma$ point lifts this
degeneracy.
As a result the solution
$(\eta_x,\eta_y,\eta_z)=(1,\epsilon,\epsilon^2)$ 
corresponding to the magnetic class $D_3(E)$ is likely. This class
allows ferromagnetism and it will have point nodes on the
$\Gamma$-centered
FS (the N point centered FS pockets will have no nodes)
\cite{gor87,sig91}.

Now consider the effect of including impurity scattering on 
$T_{c,A},T_{c,F},T_{c,E}$. It is convenient to define the following 
relaxation times
\begin{eqnarray}
(2\tau_{A})^{-1}=&\Gamma_0+4\Gamma_1+\Gamma_2 \nonumber \\
(2\tau_E)^{-1}=&\Gamma_0+\Gamma_2-2\Gamma_1 \\
(2\tau_F)^{-1}=&\Gamma_0-\Gamma_2
\end{eqnarray}
The lifetime $\tau_A$ determines the elastic mean free path
while the lifetimes $\tau_E$ and $\tau_F$ cannot be easily measured.
Solution of the linearized equations indicate 
that the transition temperatures $T_{c,i}$  are given by
\begin{equation}
\ln \frac{T_{c0,i}}{T_{c,i}}=\psi(\frac{1}{2}+\frac{\tau_{A}^{-1}-
\tau_i^{-1}}{4\pi T_{c,i}})-\psi(\frac{1}{2})
\end{equation}
where $\psi(x)$ is the digamma function.
From this expression it is clear that $T_{c,A}$ is not suppressed
by non-magnetic impurities while $T_{c,E}$ and $T_{c,F}$ are 
suppressed. Note that if $\Gamma_1=\Gamma_2=0$ (there is only
intra-pocket impurity scattering) then all three transition
temperatures   
are not suppressed by impurities. It is only inter-pocket scattering
that reduce  $T_{c,E}$ and $T_{c,F}$ from $T_{c0,E}$ and $T_{c0,F}$.   
Consequently, the suppression of $T_c$ as the impurity concentration is
increased  is not as rapid as might be expected from previous analysis
of 
unconventional single band superconductors \cite{gor85,ued85}.     
Furthermore  an impurity induced transition is possible
from the non $s$-wave superconducting states to the  
$s$-wave superconducting states when $\lambda<-4\nu-\mu$ where negative 
$\lambda$ corresponds to an
attractive intra-pocket interaction and either $\nu>0$ or
$2\nu+\mu>0$. 
The observation
of an abrupt change in the $T_c$ dependence on impurity concentration  
corresponding to
this phase transition ({\it i.e.} 
$T_c$ initially decreasing with impurity concentration in the exotic
state 
and then becoming impurity independent in the $s$-wave state)  
may help to identify such exotic superconducting states. 

The results of the above analysis may be directly applied to 
CeCo$_2$ and  predicts that due to the presence of the 
the $\Gamma$-centered FS sheet, the  low temperature specific 
heat will vary
as $T^3$ for the exotic states considered (with superconducting
classes $O(D_2)$ or $D_3(E)$). Note that
initial measurements appear more consistent with a $T^2$ behavior
however these measurements only exist down to $T/T_c \approx 0.13$
\cite{aok97} which
may not yet be sufficiently low to extract the low temperature behavior 
reliably.
For the materials LaB$_6$ and 
CeRu$_2$ the FS pockets are centered at the
X points for which the exotic superconducting state belongs to the 2D
$E_g$ 
representation which is described in detail in Ref.~\cite{ABG99}. 
Impurity 
scattering will have qualitatively the same effect  
on this state as on the exotic states examined above.  

In conclusion we have considered the effect of isotropic
non-magnetic impurity scattering  on the exotic superconducting
states that arise in a BCC lattice when the FS forms pockets centered at
the six N-points. We have found that increasing the impurity
concentration
does not suppress these non $s$-wave states as rapidly as would be
expected 
from earlier studies of impurity scattering in 
unconventional superconductors. These results are not specific to FS
pockets centered at the reciprocal lattice N points of a BCC lattice
but will apply to any degenerate set of FS pockets.

We would like to thank Z.Fisk, D. Khokhlov, J.R. Schrieffer, 
and the members
of the NHMFL condensed matter theory group seminar for
useful discussions and comments.
This work was supported by the National High Magnetic
Field Laboratory through NSF cooperative agreement 
No. DMR-9527035 and the
State of Florida.

\end{document}